\documentstyle{elsart}

\input{psfig.sty}

\def\Journal#1#2#3#4{{#1} {\bf #2}, #3 (#4)}
          
\def\NPA{{\em Nucl. Phys.} A}
\def\PLB{{\em Phys. Lett.}  B}
\def\PRL{\em Phys. Rev. Lett.}
\def\PRD{{\em Phys. Rev.} D}
\def\PRC{{\em Phys. Rev.} C}
\def\ZPC{{\em Z. Phys.} C}

\def\be{\begin{equation}}
\def\ee{\end{equation}}
\def\bea{\begin{eqnarray}}
\def\eea{\end{eqnarray}}
\def\jpsi{{\rm J}/\Psi}

\begin{document}
\begin{frontmatter}

\title{Initial State Energy Loss Dependence of 
$\jpsi$ and Drell-Yan in Relativistic Heavy 
Ion Collisions}

\author{J. L. NAGLE}

\address{Department of Physics, Columbia University\\
New York, NY 10027\\E-mail: nagle@nevis1.nevis.columbia.edu}

\author{M. J. BENNETT}

\address{Los Alamos National Laboratory\\ 
Los Alamos, NM 87545\\E-mail: mjbennett@lanl.gov}


\maketitle
\begin{abstract}
We present a Glauber-based study of $\jpsi$ and Drell-Yan yields in nucleus-nucleus 
collisions.  Using this approach, we have investigated the impact of 
energy loss by the colliding nuclei on observed yields and transverse 
momentum spectra of $\jpsi$ and Drell-Yan.  These studies permit an 
assessment of the importance of initial state energy loss in relation 
to ``anomalous'' $\jpsi$ suppression.
\end{abstract}
\end{frontmatter}

\section{Introduction}
\renewcommand{\textfraction}{0.5}

The yield of $\jpsi$ particles in relativistic heavy ion collisions is a 
subject of considerable current experimental and theoretical work.  It has 
been predicted that in the case of a phase transition to a quark-gluon plasma, 
the yield of $\jpsi$ particles is suppressed due to Debye screening~\cite{matsui}.  
The level of suppression of $\jpsi$ yields can be benchmarked in the context of 
a simple theoretical framework.  
A Glauber model is employed where each nucleon-nucleon collision is assumed to 
have an equal probability to produce a $c\overline{c}$ pair.
This $c\overline{c}$ pair may then form a quarkonium state; alternatively, the
initially produced $c\overline{c}$ pair can undergo inelastic 
interactions while traversing the nuclear medium, reducing the final $\jpsi$ yield.  
Measurements of $\jpsi$ yields from $p-A$ interactions~\cite{na38pa,e772} and 
$A-A$ interactions~\cite{na50_su,dk_quant} 
for beams of $A \le {\rm 32}$ are well described in this model by an absorption 
cross section of 6-8 mb.  
This cross section is consistent with calculated values~\cite{dk_octet} 
if the absorption occurs not on a color singlet $c\overline{c}$ pair, 
but rather on a $c\overline{c}-g$ color octet state.
However, recently the NA50 experiment has measured the $\jpsi$ yield in $Pb-Pb$
collisions at 158 $A$GeV/$c$   
as a function of the transverse energy produced in the collision in which 
the data appear to be inconsistent with a Glauber-based model including only 
the 6-8 mb absorption on nucleons~\cite{na50qm97}.  This difference is referred to as 
``anomalous'' $\jpsi$ suppression and has been interpreted by some as 
evidence for the deconfinement transition~\cite{dk_quant}.

However, without invoking deconfinement, other physical processes can and have 
been added to the above simple model.
Once the $c\overline{c}$ pair has hadronized into a specific quarkonium 
state, it may also undergo inelastic interactions in the high density medium dominated 
by mesons (often referred to as co-movers), thus producing open charm in the form
of D mesons.  Many studies indicate that these cross sections are quite 
small relative to the expected disassociation in a deconfined state with 
free quarks and gluons \cite{Mueller}.  Model extensions including interactions
with co-movers are discussed in~\cite{gavin}.  In addition, the possible
contributions of gluon shadowing~\cite{hammon} and enhanced charm 
production may play a role.  It has also been suggested that initial state energy 
loss may explain the suppression of $\jpsi$~\cite{frankel}.

In their most recent paper~\cite{na50new}, the NA50 collaboration has claimed that in 
the ratio of $\jpsi$ to Drell-Yan a ``clear onset of the anomaly is observed as a function of transverse energy.  It excludes models based on hadronic scenarios since only smooth behaviors with monotonic derivatives can be inferred from such calculations.''
This statement would imply that further investigation into explanations involving 
co-movers, initial state energy loss, {\em etc.} are not necessary.  
Although deconfinement may eventually be considered the correct explanation, we feel this 
conclusion is premature. Several studies have described some subset of the data 
reasonably well with various hadronic descriptions~\cite{gavin,hammon,frankel}. However, 
detailed studies of hadronic scenarios which compare to all of the available 
data have yet to be fully carried out, and this must be done before any real 
conclusions can be drawn.  

In this letter, we detail a study of initial state energy loss and its impact on both the 
transverse energy dependence and the transverse momentum spectra of $\jpsi$ and Drell-Yan.  In a recent report~\cite{frankel}, the inclusion of initial state
energy loss into a Glauber model was shown to match the $\jpsi$
yield measured by NA50 in $Pb-Pb$
minimum bias ({\em i.e.} averaged over all impact parameters) collisions; this agreement led to the interpretation that initial state energy loss was the source of the ``anomalous'' $\jpsi$ suppression.  Here we extend the comparison to include the
centrality dependence of the yields and $p_{t}$ spectra of both
$\jpsi$ and Drell-Yan.  By looking at the details of the entire
available data set, we hope to resolve the question of whether initial
state energy loss can explain ``anomalous'' $\jpsi$ suppression.

\section{Calculations}

\begin{figure}[t]
\psfig{figure=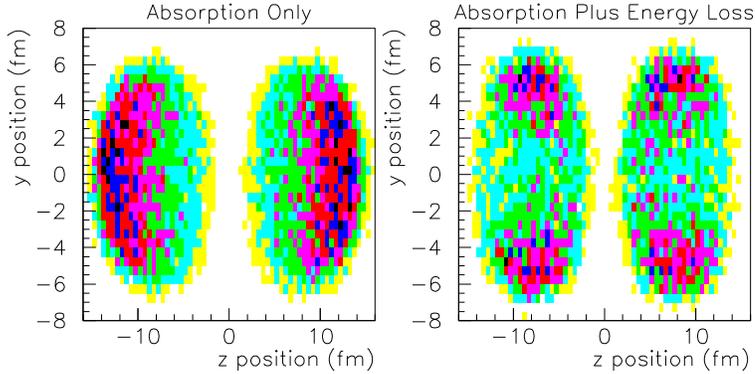,height=2.0in,bbllx=-10pt,bblly=425pt,bburx=560pt,bbury=680pt,clip=}
\caption{Side view of colliding Pb nuclei, just before interaction.  Positions are given in $z$ (along the beam axis) and $y$, transverse to the beam; a slice has been taken in $x$ to remove trivial projective density effects.  The darkness of the pixels indicates the likelihood of $\jpsi$ production and survival.  The plot on the left includes only absorption, while the plot on the right includes both absorption and initial state energy loss.    \label{fig:jpsi_geom}}
\end{figure}

In the absence of absorption and energy loss, all individual $N-N$ collisions within an $A-A$ interaction are equally likely to
produce a detected $\jpsi$.  However, due to absorption, those
nucleons on the trailing edges of the colliding nuclei are the most
likely to produce a $\jpsi$ which will survive to be detected.  This can be seen clearly in the left-hand panel of Figure~\ref{fig:jpsi_geom}, where the production position within the colliding nuclei of surviving $\jpsi$ is plotted, and only the absorption process has been included.  As
beam nucleons pass through the target nucleus, they lose energy via
inelastic interactions, so that collisions between those nucleons on
the trailing edges of the nuclei, where the geometry is most favorable
for a produced $\jpsi$ to evade absorption, have considerably less
than the full beam energy.  Since $\jpsi$ production has a very steep
energy dependence~\cite{schuler}, these collisions are the least
likely to produce a $\jpsi$.  Thus, nucleon energy loss will certainly
reduce the observed yield of $\jpsi$.  This can be seen in the right-hand panel of Figure~\ref{fig:jpsi_geom}, where the production position within the colliding nuclei of surviving $\jpsi$ is plotted, and both initial state energy loss and absorption has been included.  Moreover, the ratio of $\jpsi$
to Drell-Yan, which is often used to gauge $\jpsi$ suppression, could
have a complicated centrality dependence, since Drell-Yan does not
suffer from the geometrical predisposition caused by absorption.
Finally, energy loss will also affect the $p_{t}$ spectrum of
$\jpsi$--the trailing edge collisions which are most affected by
energy loss are those which, via the Cronin effect~\cite{cronin},
produce $\jpsi$ with the highest mean $p_{t}$.

In order to study initial state energy loss, we have constructed a 
Glauber model of nuclear collisions.  We will describe the model
briefly here; more details are available elsewhere~\cite{ourprc}.
Nucleons are distributed using a Woods-Saxon parameterization, and
interact with a nucleon-nucleon cross section of 30 mb.  As nucleons
undergo interactions, they lose energy.  Various models of energy loss
are reasonably consistent with measured proton spectra in $p-A$
collisions; we have utilized a parameterization where nucleons lose a
constant fraction of their momentum in each interaction.  In order to
match measured data, the momentum fraction lost per interaction would
be $\sim$40\%.  However, most of this energy loss occurs via soft
interactions, with a time scale of a few fm/$c$.  At SPS energies, the
colliding nuclei cross in $\sim$0.1 fm/$c$; thus, only a fraction of
the time-integrated total energy loss is applicable to hard
interactions.  Our approach is to treat the applicable fraction of
total energy loss as a variable parameter.  The values we have chosen
are 5\%, 10\% and 15\% momentum loss per collision, to be compared with
the 40\% loss realized as the time between collisions approaches $\infty$.
By counting the number of prior collisions for each nucleon, a
center-of-mass energy can be calculated for each $N-N$ interaction.
This energy is used to calculate the relative probability that a
$\jpsi$ or Drell-Yan pair will be produced, using the Schuler
parameterization~\cite{schuler} for the $\jpsi$ energy dependence and
``tau scaling''~\cite{tau} for Drell-Yan energy dependence.

Produced $\jpsi$ are taken to be at rest in the $N-N$ center-of-mass
frame, such that the survival probability is a function of the path
length through nuclear material which the $\jpsi$ must traverse, the
nuclear density and the breakup cross section.  We have utilized a
breakup cross section of 7.1 mb, which is $\sim$15\% higher than the
value of 6.2 $\pm$ 0.7 mb reported by NA50~\cite{na50qm97}.  The NA50
value is calculated by fitting the $\jpsi$ to Drell-Yan ratio as an
exponential function of $L$, the mean path length through the nuclear
medium, for various centrality bins in $p-A$ and $A-A$ collisions.
Due to absorption, all possible path lengths do not contribute equally
to the {\em actual} mean path length for surviving $\jpsi$ in a given
centrality bin; thus, as we have shown elsewhere~\cite{ourprc}, a
simple linear average over path lengths systematically underestimates
the absorption cross section.

\begin{figure}[t]
\psfig{figure=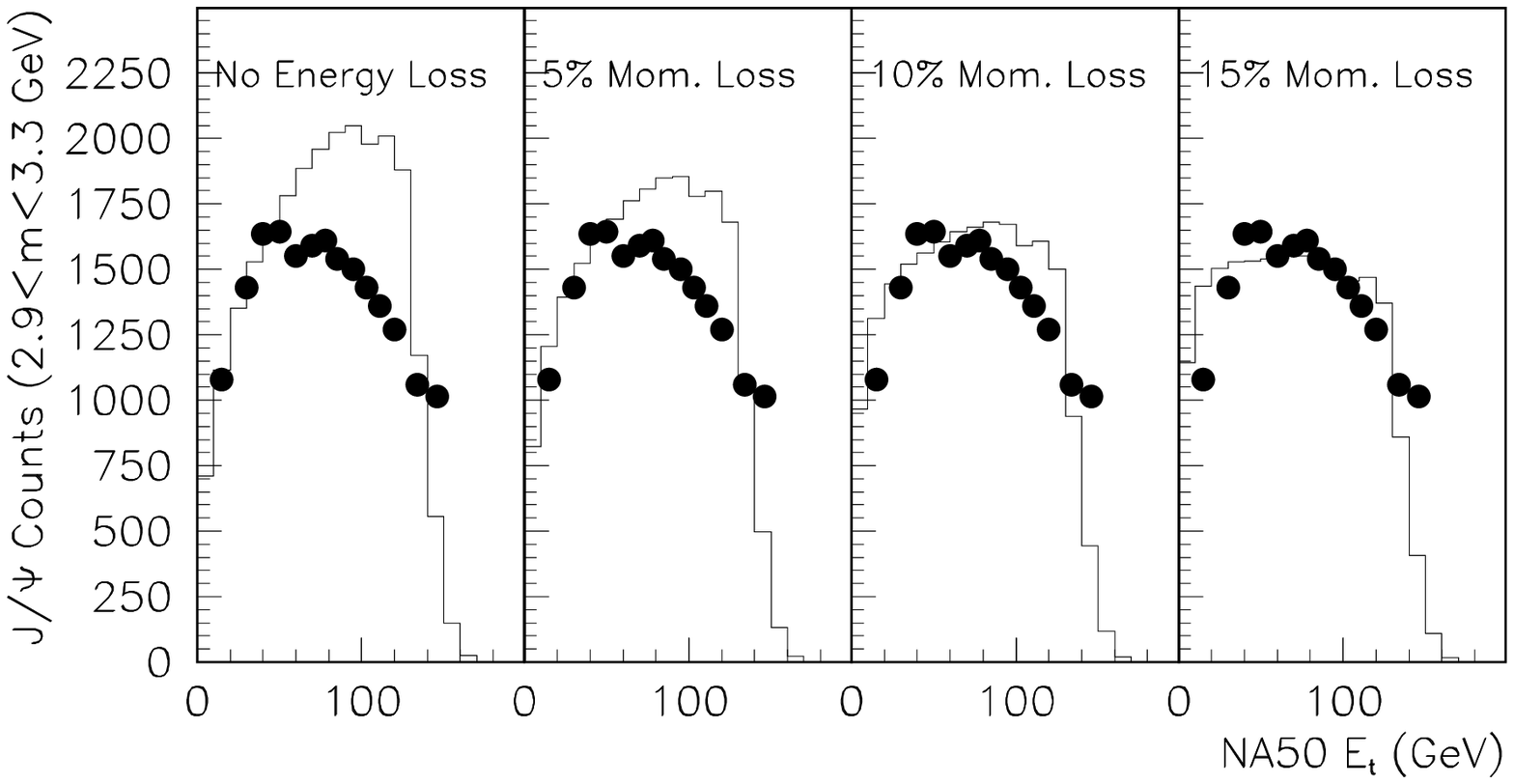,height=2.0in,bbllx=-75pt,bblly=375pt,bburx=560pt,bbury=680pt,clip=}
\caption{Comparison of Glauber model calculations (line) to $\jpsi$
yields from $Pb-Pb$ collisions as measured by NA50 (points), plotted
as a function of $E_{t}$.  The various panels show different values of
initial state energy loss in terms of momentum fraction lost per $N-N$
collision.   \label{fig:jpsi_yield}}
\end{figure}

\begin{figure}[t]
\psfig{figure=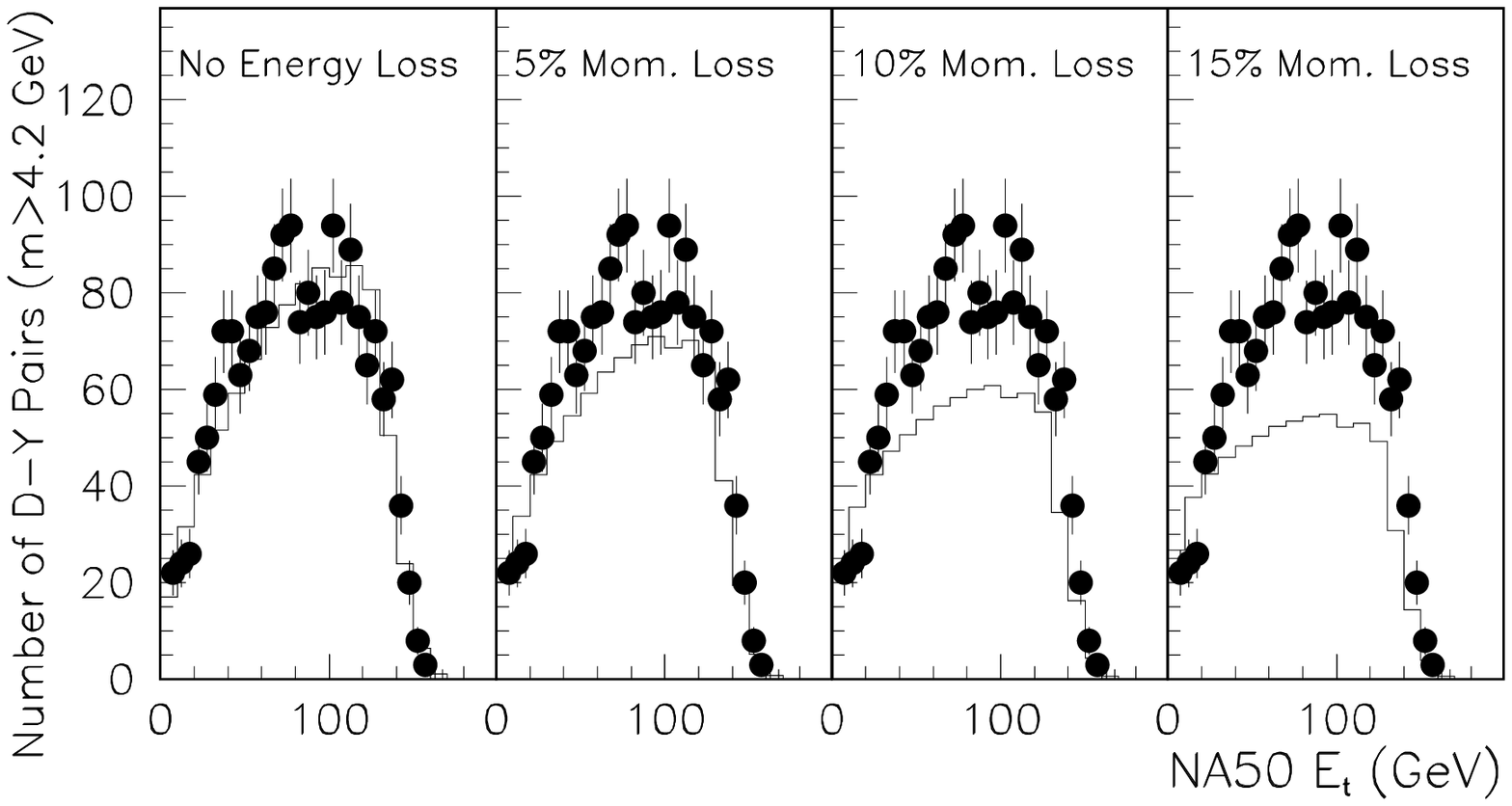,height=2.0in,bbllx=-75pt,bblly=375pt,bburx=560pt,bbury=680pt,clip=}
\caption{Comparison of Glauber model calculations (line) to Drell-Yan
yields from $Pb-Pb$ collisions as measured by NA50 (points), plotted
as a function of $E_{t}$.  The various panels show different values of
initial state energy loss in terms of momentum fraction lost per $N-N$
collision. \label{fig:dy_yield}}
\end{figure}

Shown in Figures~\ref{fig:jpsi_yield} and~\ref{fig:dy_yield} are the
calculated yields of  $\jpsi$ and Drell-Yan, respectively, from
$Pb-Pb$ collisions, plotted as a function of transverse energy
$E_{t}$, and compared to the NA50 measured values~\cite{na50qm97}.  We
have simulated the NA50 $E_{t}$ bins, assuming $E_{t}$ scales as the
number of wounded nucleons and smearing the calculated values by the
NA50 resolution of 94\%/$\sqrt{E_{t}}$~\cite{na50_resolution}.  For each figure, the yield
without energy loss is shown in the leftmost panel, as well as that
for our nominal choices of 5\%, 10\% and 15\% momentum loss per $N-N$
collision.  Our model does not predict absolute yields, so the
normalization for each curve has been chosen so as to best match the
NA50 data in the lowest $E_{t}$ bins.  Clearly, the $\jpsi$ yield
deviates significantly from the prediction with no energy loss;
although not shown here, the discrepancy is significant even if one
increases the $\jpsi$ breakup cross section as high as 9 mb.  This plot
shows, in the simplest way, the additional suppression seen in the $Pb-Pb$ data as compared to expectations based on lighter systems.  In the following panels of Figure~\ref{fig:jpsi_yield}, it can be seen that as
energy loss is invoked, the prediction comes closer to the data, until
for the maximum value of energy loss we consider, the prediction
matches the data relatively well.  

However, the scenario for Drell-Yan is considerably different, as shown in Figure~\ref{fig:dy_yield}.  
The prediction for no energy loss matches
the data very well, while with just 5\% momentum loss per collision,
the prediction deviates significantly from most of the data points.
For the maximum value of energy loss, which is necessary to match the
$\jpsi$ yields, the prediction does not come close to matching the
data.  Thus, there is an inconsistency--the model can be forced to
match the $\jpsi$ data by invoking a fairly large amount of energy
loss, but this same value of energy loss causes the model to
substantially underestimate the Drell-Yan yields.  However, it is
possible that this inconsistency could be explained away if the energy
loss of gluons, which may dominate $\jpsi$ production, is different from
that of quarks and antiquarks, the mutual annihilation of which lead
to Drell-Yan production; whether this is the case remains an open
question.  It should be noted that precision studies
of how energy loss affects Drell-Yan in $p-A$ collisions are also 
being done~\cite{leitch,kapusta}.  

Another way to consider the effects of energy loss is to look at the
$p_{t}$ spectra of $\jpsi$.  The $\langle p_{t}^{2} \rangle$ of
$\jpsi$ from $p-A$ collisions has been observed to be larger than that
from $p-p$ collisions.  This increased $p_{t}$ is
understood~\cite{cronin} to come from an increase in the intrinsic
transverse momentum of the partons in the colliding nucleons as a
result of prior interactions.  This mechanism, referred to as the
``Cronin effect'', has been phenomenologically described by
\begin{equation}
 \langle p_{t}^{2} \rangle_{N} = \langle p_{t}^{2} \rangle_{pp} + N
 \Delta p_{t}^{2},
\end{equation}
in which the $\langle p_{t}^{2} \rangle$ of $\jpsi$ produced in a
nucleon-nucleon collision where the colliding partners had a total of
$N$ prior interactions is given by the sum of the $\langle p_{t}^{2}
\rangle$ value from $p-p$ collisions plus $N$ times $\Delta
p_{t}^{2}$, the change in $\langle p_{t}^{2} \rangle$ per collision.
Thus, $\jpsi$ with the highest mean $p_{t}$ come from the latest collisions, and
are the most sensitive to effects coming from nucleon energy loss.

For $\jpsi$, a value of $ \langle p_{t}^{2} \rangle_{pp}$ = 1.23 $\pm$ 0.05 GeV$^{2}$ was measured by NA3\cite{na3_jpsi} at a beam momentum of 200 $A$GeV/$c$;
however, before we can use this value, we must correct for the beam energy, since the SPS $Pb$ beams are at 160 $A$GeV/$c$.
Measured $ \langle p_{t}^{2} \rangle$ for Drell-Yan from pion and proton induced reactions on nuclei have been shown\cite{na3_dy} to scale linearly with $s$, the square of the center-of-mass energy, with similar slopes for the two incident particle species.
Measurements of $ \langle p_{t}^{2} \rangle$ for $\jpsi$ from proton induced reactions are scarce, but if one combines data from both pion and proton induced reactions\cite{na3_jpsi}, a linear scaling with $s$ fits the data reasonably well.  Using this slope, we estimate the value of $ \langle p_{t}^{2} \rangle_{pp}$ for interactions at 160 $A$GeV/$c$ to be 1.13 GeV$^{2}$. 
A recent study\cite{dk_pt}, combining $\jpsi$ data from $p-A$ and $A-A$
interactions at 200 $A$GeV/$c$ and using the measured value of $ \langle
p_{t}^{2} \rangle_{pp}$ given above, performed a fit to determine a value for $\Delta p_{t}^{2}$ of 0.125
GeV$^{2}$.

Using these parameters, we have implemented the Cronin effect in our model. Transverse momentum distributions are taken to follow the usual form of $\frac{dN}{dm_{t}} \propto m_{t} \exp{(-\alpha m_{t})}$.  The prescription for the Cronin effect characterizes the transverse momentum after $N$ collisions in terms of $\langle p_{t}^{2} \rangle_{N}$, which is related to a slope parameter $\alpha_{N}$ by
\begin{equation}
\langle p_{t}^{2} \rangle_{N} = \frac{2}{m \alpha_{N} + 1}\left[
\frac{3}{\alpha_{N}^{2}} + \frac{3m}{\alpha_{N}} + m^{2}\right]
\label{eqn:slope}
\end{equation}
for a particle of mass $m$.  In practice, we wish to convert $\langle p_{t}^{2} \rangle_{N}$ to $\alpha_{N}$; over the $\langle p_{t}^{2} \rangle$ range of interest and for the $\jpsi$ mass, the inverse of Equation~\ref{eqn:slope} is well approximated by a power law,
$\alpha_{N} \approx 6.68 \times \langle p_{t}^{2} \rangle_{N}^{-0.855}$.
In the course of the Glauber calculation, a $p_{t}$ value is chosen from the appropriate distribution (based on the number of prior $N-N$ collisions for the interacting nucleons) for each produced $\jpsi$.  For those $\jpsi$ which evade absorption, a running value for $\langle p_{t}^{2} \rangle$ is tabulated as a function of the total $E_{t}$ of the collision.      

\begin{figure}[t]
\psfig{figure=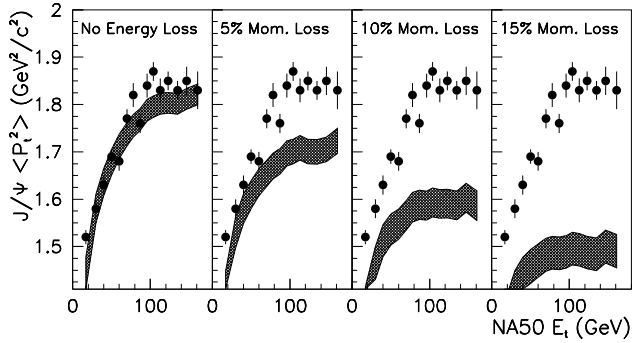,height=2.0in,bbllx=-75pt,bblly=375pt,bburx=560pt,bbury=680pt,clip=}
\caption{Glauber model calculations (shaded band) of $\jpsi$ $ \langle
p_{t}^{2} \rangle$ from $Pb-Pb$ collisions, compared to NA50 measured
data (points), plotted as a function of $E_{t}$.  The width of the
shaded band indicates the uncertainty in the measured value of
$\langle p_{t}^{2} \rangle$ from $pp$ collisions, which has been
energy scaled (see text).  The various panels show different values of
initial state energy loss in terms of momentum fraction lost per $N-N$
collision.\label{fig:jpsi_pt2}}
\end{figure}

The prediction for $\jpsi$ $\langle p_{t}^{2} \rangle$ from $Pb-Pb$
collisions is compared to the NA50 data in Figure~\ref{fig:jpsi_pt2}.
The prediction is shown as a band of values, corresponding to the
uncertainty in the scaled value of $ \langle p_{t}^{2} \rangle_{pp}$.
The prediction with no energy loss matches the NA50 data quite well;
this result is at odds with other recent studies~\cite{na50qm97}, in
which it has been claimed that a plasma was required in order to match
the NA50 data.  However, these calculations did not include a beam
energy rescaling of the value for $ \langle p_{t}^{2} \rangle_{pp}$.
There is some uncertainty in the scaling we have implemented, so that
the question of matching the NA50 data is still open; however, given
the normalization uncertainty involved, it seems premature to rule out
a normal hadronic description of the NA50 $\langle p_{t}^{2} \rangle$ data.

It is clear, however, that the inclusion of energy loss causes the
prediction to deviate severely from the data.  For a value of 15\%
momentum loss per collision, which gave the best match to the $\jpsi$
yields, the prediction for  $\langle p_{t}^{2} \rangle$ is in severe
disagreement with the data.  Thus, again we have an
inconsistency; in this case, a single value of energy loss cannot
describe both the $\jpsi$ yields and the $\langle p_{t}^{2} \rangle$
data.  Since this discrepancy is between two aspects of the energy
dependence of $\jpsi$, it is not so easily dismissed.

\section{Conclusions}

In summary, we have performed an evaluation of the importance of
initial state energy loss with respect to $\jpsi$ suppression.  Within
the uncertainty in normalization, the $\jpsi$ $\langle p_{t}^{2}
\rangle$ spectrum seems consistent with a normal hadronic scenario.
The addition of energy loss can cause the model prediction to fit the
$\jpsi$ yields; however, a single value of energy loss per collision
cannot simultaneously match both the $\jpsi$ and Drell-Yan yields, nor
can it simultaneously match both the $\jpsi$ yields and the $\jpsi$
$\langle p_{t}^{2} \rangle$ data.  This result suggests that, contrary to the proposal made elsewhere~\cite{frankel} based on minimum bias $\jpsi$ yields only,
``anomalous'' $\jpsi$ suppression cannot be explained by initial state
energy loss.  Clearly the simplest hadronic model does not match the $Pb-Pb$ $\jpsi$ yields, so that some other mechanism must be causing the
increased suppression.  However, before one can either claim or rule out any source of this effect, whether ``normal'' hadronic or otherwise, a systematic comparison to all of the data, as we have attempted to do here, must be performed.


\section{Acknowledgements}
This work has been supported by the grant from the U.S. Department of
Energy to Columbia University DE-FG02-86ER40281 and the Contract to 
Los Alamos National Laboratory W-7405-ENG-36.


\begin{thebibliography}{99}

\bibitem{matsui}T. Matsui and J. Satz, \Journal{\PLB}{220}{416}{1986}.
\bibitem{na38pa}C. Louren\c{c}o, \Journal{\NPA}{610}{552c}{1996}.
\bibitem{e772}M.J. Leitch {\em et al.},\Journal{\PRD}{52}{4251}{1995}.
\bibitem{na50_su}M.C. Abreu {\em et al.},
\Journal{\PLB}{449}{128}{1999}.
\bibitem{dk_quant}D. Kharzeev {\em et al.},
\Journal{\ZPC}{74}{307}{1997}.
\bibitem{dk_octet}D. Kharzeev and H. Satz,
\Journal{\PLB}{366}{316}{1996}.
\bibitem{na50qm97}L. Ramello for the NA50 Collaboration,
\Journal{\NPA}{638}{261}{1998}.
\bibitem{Mueller}S.G. Matinian and B. Muller, 
\Journal{\PRC}{58}{2994}{1998}.
\bibitem{gavin}S. Gavin and R. Vogt, \Journal{\PRL}{78}{1006}{1997}.
\bibitem{hammon}N. Hammon {\em et al.}, \Journal{\PRC}{59}{2744}{1999}.
\bibitem{frankel}S. Frankel and W. Frati,
\Journal{\PLB}{441}{425}{1998}.
\bibitem{na50new}M.C. Abreu {\em et al.},
\Journal{\PLB}{450}{456}{1999}.
\bibitem{schuler}G.A. Schuler and R. Vogt,
\Journal{\PLB}{387}{181}{1996}.
\bibitem{cronin}S. Gavin and M. Gyulassy,
\Journal{\PLB}{214}{241}{1988}.
\bibitem{tau}T.-M. Yan, hep-th/9810268.
\bibitem{ourprc}M.J. Bennett and J.L. Nagle,
\Journal{\PRC}{59}{2713}{1999}.
\bibitem{na50_resolution}S. Beole, Looking for Quark Gluon Plasma in $Pb-Pb$ Collisions at 158 GeV/$c$, Universita Degli Studi Di Torino, Ph.D. Thesis (1998).
\bibitem{leitch}M.A. Vasilev {\em et al.} for FNAL E866 Collaboration, hep-ex/9906010.
\bibitem{kapusta}C. Gale, S. Jeon, J. Kapusta, 
\Journal{\PRL}{82}{1636}{1999}.
\bibitem{na3_jpsi}J. Badier {\em et al.} (NA3 Collaboration),
\Journal{\ZPC}{20}{101}{1983}.
\bibitem{na3_dy}J. Badier {\em et al.} (NA3 Collaboration),
\Journal{\PLB}{117}{372}{1982}.
\bibitem{dk_pt}D. Kharzeev, M. Nardi and H. Satz,
\Journal{\PLB}{405}{14}{1997}.

\end{thebibliography}
\end{document}